\documentclass[twocolumn,runningheads]{svjour2}
\smartqed  
\usepackage{graphicx}
\usepackage{mathptmx}   

\usepackage[hyperindex,breaklinks]{hyperref}

\usepackage[authoryear]{natbib}
\newcommand{\noprintlabel}{}
\newcommand{\sign}{{\rm sign \,}}
\newcommand{\eye}{{\rm i}}
\newcommand{\ie}{i.e.}
\newcommand{\eg}{e.g.}
\newcommand{\cf}{cf.}
\newcommand{\etal}{et al.}
\newcommand{\xf}{x_{\rm f}}
\newcommand{\xs}{x_{\rm s}}
\newcommand{\xih}{\xi_{\rm h}}

\def\biblab #1{\ifx\noprintlabel\undefined{\bf [#1]}\fi}
\newcommand{\CI}{{\cal I}}
\newcommand{\CK}{{\cal K}}
\newcommand{\CO}{{\cal O}}

\newcommand{\dd}{{\rm d}}

\journalname{Astrophysics and Space Science (CoRoT/ESTA Volume)}
\begin{document}

\title{ADIPLS -- the Aarhus adiabatic oscillation package
}


\author{J{\o}rgen Christensen-Dalsgaard
}


\institute{J{\o}rgen Christensen-Dalsgaard \at
              Institut for Fysik og Astronomi, og Dansk AsteroSeismisk Center \\
              Bygning 1520 \\
              Aarhus Universitet \\
              DK-8000 Aarhus C \\
              Denmark \\
              Tel.: +45 89 42 36 14\\
              Fax: +45 86 12 07 40\\
              \email{jcd@phys.au.dk}           
}

\date{Received: date / Accepted: date}

\maketitle
\begin{abstract}
Development of the Aarhus adiabatic pulsation code started around 1978.
Although the main features have been stable for more than a decade,
development of the code is continuing, concerning numerical properties
and output.
The code has been provided as a generally available package and
has seen substantial use at a number of installations.
Further development of the package, including bringing the documentation
closer to being up to date, is planned as part of the HELAS Coordination
Action.
\end{abstract}

\section{Introduction}\label{intro}

The goal of the development of the code was to have a simple and efficient 
tool for the computation of adiabatic oscillation frequencies and 
eigenfunctions for general stellar models, emphasizing also the accuracy
of the results.
Not surprisingly, given the long development period, the simplicity 
is now less evident.
However, the code offers considerable flexibility in the choice of
integration method as well as ability to determine all frequencies of
a given model, in a given range of degree and frequency.

The choice of variables describing the equilibrium model and oscillations
was to a large extent inspired by \citet{Dziemb1971}.
As discussed in Section~\ref{sec:eqmodel} the equilibrium model is
defined in terms of a minimal set of dimensionless variables, as
well as by mass and radius of the model.

Fairly extensive documentation of the code,
on which the present paper in part is based, is provided with the distribution
package%
\footnote{The package is available at \\
{\tt http://astro.phys.au.dk/$\sim$jcd/adipack.n}}.
\citet{Christ1991} provided an extensive review of adiabatic stellar oscillations,
emphasizing applications to helioseismology,
and discussed many aspects and tests of the Aarhus package,
whereas \citet{Christ1994} carried out careful tests and comparisons of
results on polytropic models;
this includes extensive tables of frequencies which can be used for comparison
with other codes.

\section{Equations and numerical scheme}\label{sec:eqnum}

\subsection{Equilibrium model}
\label{sec:eqmodel}

The equilibrium model is defined in terms of the following dimensionless
variables:
\begin{eqnarray}
x &\equiv& r / R \; , \nonumber \\
A_1 &\equiv& q / x^3  , \qquad \hbox{\rm where } q = m / M \; , \nonumber \\
A_2 &=& V_g \equiv -  {1 \over \Gamma_1} {\dd \ln p  \over \dd \ln r}
= {G m \rho   \over \Gamma_1 p r} \; , \nonumber \\
A_3 &\equiv& \Gamma_1 \; , \nonumber \\
A_4 &=& A \equiv {1 \over \Gamma_1} {\dd \ln p \over \dd \ln r} -
{\dd \ln \rho   \over \dd \ln r} \; , \nonumber \\
A_5 &=& U \equiv {4 \pi \rho r^3  \over m } \; .
\label{eq:fivea}
\end{eqnarray}
Here $r$ is distance to the centre, $m$ is the mass interior to $r$,
$R$ is the photospheric radius of the model and $M$ is its mass;
also, $G$ is the gravitational constant,
$p$ is pressure, $\rho$ is density, and
$\Gamma_1 = (\partial \ln p / \partial \ln \rho)_{\rm ad}$,
the derivative being at constant specific entropy.
In addition, the model file defines $M$ and $R$, as well as central pressure
and density, in dimensional units, and scaled second derivatives of $p$
and $\rho$ at the centre (required from the expansions in the central
boundary condition); 
finally, for models with vanishing surface pressure, assuming a polytropic
relation between $p$ and $\rho$ in the near-surface region, the polytropic
index is specified.

The following relations between the variables defined here and
more ``physical'' variables are often useful:
\begin{eqnarray}
p &=& {G M^2  \over 4 \pi R^4}  {x^2 A_1^2 A_5 \over A_2 A_3} \; , \quad
{\dd p \over \dd r }= - {G M^2  \over 4 \pi R^5} x A_1^2 A_5 \; , \nonumber \\
\rho &=& {M \over 4 \pi R^3}  A_1 A_5 \; .
\end{eqnarray}
We may also express the characteristic frequencies for adiabatic
oscillations in terms of these variables. Thus if $N$ is the
buoyancy frequency, $S_l$ is the Lamb frequency at degree $l$
and $\omega_{\rm a}$ is the acoustical cut-off frequency 
for an isothermal atmosphere, we have
\begin{eqnarray}
\label{eq:buoy}
N^2 &=& {G M \over R^3} {\hat N}^2 =  {G M \over R^3}  A_1 A_4 \; , \\
\label{eq:lamb}
S_l^2 &=& {l ( l + 1) c^2  \over r^2}
= {G M \over R^3} {\hat S}_l^2 =
{G M \over R^3} {l ( l + 1) A_1  \over  A_2} \; , \\
\label{eq:cutoff}
\omega_{\rm a}^2 &=& {c^2  \over 4 H_p^2} 
= {G M \over R^3} \hat \omega_{\rm a}^2
= {1 \over 4 } {G M \over R^3}  A_1 A_2 A_3^2 \; ,
\end{eqnarray}
where $c$ is the adiabatic sound speed, and $H_p = p / ( g \rho )$ is
the pressure scale height, $g$ being the gravitational acceleration.
Finally it may be noted that the squared sound speed is given by
\begin{equation}
c^2 = {G M \over R } \hat c^2 
= {G M \over R }x^2 {A_1  \over A_2} \; .
\label{eq:E5.7}
\end{equation}
These equations also define the dimensionless characteristic frequencies
$\hat N$, $\hat S_l$ and $\hat \omega_{\rm a}$ as well as the dimensionless
sound speed $\hat c$, which are often useful.

\subsection{Formulation of the equations}\label{sec:eq}

As is well known the displacement vector of nonradial (spheroidal) modes
can be written in terms of polar coordinates $(r, \theta, \phi)$ as
\begin{eqnarray}
\vec{\delta r} &=& {\rm Re} \mbox{\huge$\left\{ \left[\right.\right.$}
\xi_r (r) Y_l^m ( \theta ,
\phi ) {{\bf a} }_r + \\ 
&& \left. \left. \xih (r) \left( { \partial Y_l^m
\over \partial \theta } {{\bf a} }_{\theta} + {1 \over \sin  \theta}
{\partial Y_l^m  \over \partial \phi} {{\bf a} }_{\phi} \right)
\right] \exp (- \eye \omega t ) \right\} \; . \nonumber
\end{eqnarray}
Here $Y_l^m(\theta, \phi) = c_{lm} P_l^m(\cos \theta) \exp( \eye m \phi)$
is a spherical harmonic of degree $l$ and
azimuthal order $m$, $\theta$ being co-latitude and $\phi$ longitude;
$P_l^m(x)$ is an associated Legendre function, and $c_{lm}$ is a
suitable normalization constant.
Also, ${{\bf a}}_r$, ${{\bf a}}_{\theta}$, and
${{\bf a}}_{\phi}$ are unit vectors in the
$r$, $\theta$, and $\phi$ directions.
Finally, $t$ is time and $\omega$ is the angular frequency of the mode.
Similarly, e.g., the Eulerian perturbation to pressure may be written%
\footnote{I do not here distinguish between the full perturbation and the
radial amplitude function.}
\begin{equation}
p' (r, \theta, \phi, t) =
{\rm Re} \left[ p'(r) Y_l^m ( \theta , \phi ) 
\exp (- \eye \omega t  ) \right] \; .
\label{eq:E2.2}
\end{equation}
As the oscillations are adiabatic (and only conservative boundary
conditions are considered) $\omega$ is real, and the amplitude functions
$\xi_r (r)$, $\xih (r)$, $p' (r)$, etc. can be chosen to be real.

The equations of adiabatic stellar oscillations, in the nonradial case,
are expressed in terms of the following variables:%
\footnote{The somewhat peculiar choice of $y_3$, $y_4$ results from the
earlier use of an unconventional sign convention for $\Phi'$;
now, as usual, $\Phi'$ is defined such that the perturbed Poisson equation
has the form $\nabla^2 \Phi' = 4 \pi G \rho'$,
where $\rho'$ is the Eulerian density perturbation.}
\begin{eqnarray}
y_1 &=& {\xi_r  \over R } \; , \nonumber \\
y_2 &=& x  \left( {p^{\prime}  \over \rho } +  \Phi^{\prime} \right)
{l ( l + 1)  \over \omega^2 r^2} = {l ( l + 1)  \over R }  \xih \; , 
\nonumber \\
y_3 &=&  - x  {\Phi^{\prime}  \over gr } \; , \nonumber \\
y_4  &=& x^2  {\dd \over \dd x }\left( {y_3   \over x }\right) \; . 
\label{eq:E2.5a}
\end{eqnarray}
Here $\Phi'$ is the perturbation to the gravitational potential.
Also, we introduce the dimensionless frequency $\sigma$ by
\begin{equation}
\omega^2 = {G M \over R^3} \sigma^2 \; ,
\end{equation}
corresponding to Eqs \ref{eq:buoy} -- \ref{eq:cutoff}.
These quantities satisfy the following equations:
\begin{eqnarray}
\label{eq:EA.1}
x  {\dd y_1  \over \dd x } &=& (V_g - 2 )  y_1 +
\left( 1 - {V_g  \over \eta }\right)  y_2 - V_g  y_3 \; , \\
\label{eq:EA.2}
x {\dd y_2  \over \dd x } &=& [ l ( l + 1) - \eta A ]  y_1
+ (A - 1 )  y_2 + \eta A  y_3 \; , \\
\label{eq:EA.3}
x  {\dd y_3  \over \dd x } &=& y_3 + y_4 \; , \\
\label{eq:EA.4}
x  {\dd y_4  \over \dd x } &=& - A U  y_1  
- U {V_g   \over \eta } y_2 \\
&& + [ l ( l + 1) + U(A - 2 ) +  U V_g ]  y_3 + 2(1 - U )  y_4 \; . \nonumber
\end{eqnarray}
Here $\eta = l ( l + 1) g / ( \omega^2 r ) = l(l+1) A_1/\sigma^2$,
and the notation is otherwise
as defined in Eq.~\ref{eq:fivea}. 
In the \citet{Cowlin1941} approximation,
where the perturbation to the gravitational potential is neglected,
the terms in $y_3$ are neglected in Eqs~\ref{eq:EA.1} and \ref{eq:EA.2}
and Eqs~\ref{eq:EA.3} and \ref{eq:EA.4} are not used.

The dependent variables $y_i $ in the nonradial case
have been chosen in such a way that
for $l > 0$ they all vary as $x^{l - 1}$ for $x \rightarrow 0$.
For large $l$
a considerable (and fundamentally unnecessary) computational effort would
be needed to represent this variation sufficiently accurately with, {\eg},
a finite difference technique, if these variables were to be used in
the numerical integration. Instead I introduce a new set of dependent
variables by
\begin{equation}
{\hat y}_i = x^{- l +1}  y_i , \qquad i  = 1, 2, 3, 4 \; .
\label{eq:E2.6}
\end{equation}
These variables are then $O(1)$ in $x$ near the centre.
They are used in the region where the variation in the
$y_i$ is dominated by the $x^{l - 1}$ behaviour, for $x < x_{\rm ev}$,
say, where $x_{\rm ev}$ is determined on the basis of the asymptotic properties
of the solution.
This transformation permits calculating modes of arbitrarily high degree
in a complete model.

For radial oscillations only $y_1$ and $y_2$ are used, where
$y_1$ is defined as above, and 
\begin{equation}
y_2 = {p^{\prime}  \over \omega^2 R^2 \rho} \; .
\label{eq:E2.5b}
\end{equation}
Here the equations become
\begin{eqnarray}
x  {\dd y_1  \over \dd x } &=& (V_g - 2 )  y_1 -
V_g {\sigma^2 x^2   \over q } y_2 \; , \\
\label{eq:EA.5}
x {\dd y_2  \over \dd x } &=& \left[ x - {q \over \sigma^2 x^2 }
(A - U) \right] y_1 + A  y_2 \; .
\label{eq:EA.6}
\end{eqnarray}

The equations are solved on the interval $[x_1, \xs]$ in $x$.
Here, in the most common case involving a complete stellar model
$x_1 = \epsilon$, where $\epsilon$ is a suitably small number such that the
series expansion around $x = 0$ is sufficiently accurate;
however, the code can also deal with envelope models with arbitrary $x_1$,
typically imposing $\xi_r = 0$ at the bottom of the envelope.
The outermost point is defined by $\xs = R_{\rm s}/R$ where $R_{\rm s}$
is the surface radius, including the atmosphere; thus, typically, $\xs > 1$.

\subsection{Boundary conditions}

The centre of the star, $r = 0$, is obviously a singular point of 
the equations.
As discussed, e.g., by \citet{Christ1974} boundary conditions at this
point are obtained from a series expansion, in the present code to 
second significant order.
In the general case this defines two conditions at the innermost non-zero
point in the model.
For radial oscillations, or nonradial oscillations in the Cowling 
approximation, one condition is obtained.
The surface in a realistic model is typically defined at a suitable
point in the stellar atmosphere, with non-zero pressure and density.
Here the simple condition of vanishing Lagrangian pressure perturbation
is implemented and sometimes used.
However, more commonly a condition between pressure perturbation and
displacement is established by matching continuously to the solution
in an isothermal atmosphere extending continuously from the uppermost
point in the model.%
\footnote{Note that since the frequency, and other variables,
are taken to be real
this can only be applied for frequencies below the acoustical cut-off
frequency in the isothermal extension.}
A very similar condition was presented by \citet{Unno1989}.
In addition, in the full nonradial case a condition is obtained from
the continuous match of $\Phi'$ and its derivative to the vacuum solution
outside the star.

In full polytropic models, or other models with vanishing surface pressure,
the surface is also a singular point.
In this case a boundary condition at the outermost non-singular point is
obtained from a series expansion, assuming a near-surface polytropic 
behaviour \citep[see][for details]{Christ1994}.

The code also has the option of considering truncated (e.g., envelope) models
although at the moment only in the Cowling approximation or for radial 
oscillations.
In this case the innermost boundary condition is typically the vanishing of
the radial displacement $\xi_r$ although other options are available.

\subsection{Numerical scheme}

The numerical problem can be formulated generally as that of solving
\begin{equation}
{\dd y_i  \over \dd x} = \sum_{j=1}^I a_{ij} (x) y_j (x) \; , \qquad
\hbox{\rm for }  i = 1, \ldots , I \; ,
\label{eq:E3.1}
\end{equation}
with the boundary conditions
\begin{equation}
\sum_{j=1}^I b_{ij} y_j (x_1 ) = 0 \; , \qquad
\hbox{\rm for }  i = 1, \ldots \; ,  I/2 \; ,
\label{eq:E3.2}
\end{equation}
\begin{equation}
\sum_{j=1}^I c_{ij} y_j (\xs ) = 0 \; , \qquad
\hbox{\rm for }  i =  1 \; , \ldots,  I/2 \; .
\label{eq:E3.3}
\end{equation}
Here the order $I$ of the system is 4 for the full nonradial case, and
2 for radial oscillations or nonradial oscillations in the Cowling 
approximation. This system only allows non-trivial solutions for
selected values of $\sigma^2$ which is thus an eigenvalue of
the problem.

The programme permits solving these equations with two basically different
techniques, each with some variants. The first is a shooting method,
where solutions satisfying the boundary conditions are integrated 
separately from the inner and outer boundary, and the eigenvalue
is found by matching these solutions at a suitable inner fitting point $\xf$.
The second technique is to solve the equations together with a 
normalization condition and all
boundary conditions using a relaxation technique; the eigenvalue is
then found by requiring continuity of one of the eigenfunctions at
an interior matching point.

For simplicity I do not distinguish between $\hat y_i$ and
$y_i$ ({\cf} Section~\ref{sec:eq}) in this section.
It is implicitly understood
that the dependent variable (which is denoted $y_i$) is
$\hat y_i$ for $x < x_{\rm ev}$ and $y_i$ for $x \ge x_{\rm ev}$.
The numerical treatment of the transition between $\hat y_i$ and 
$y_i$ has required a little care in the coding.

\subsection{The shooting method} 

It is convenient here to distinguish between $I$ = 2 and $I$ = 4.
For $I$ = 2 the differential Eqs~\ref{eq:E3.1} have a unique (apart
from normalization) solution $y_i^{\rm (i)}$ satisfying the
inner boundary conditions \ref{eq:E3.2}, 
and a unique solution $y_i^{\rm (o)}$
satisfying the outer boundary conditions \ref{eq:E3.3}. These are obtained
by numerical integration of the equations. 
The final solution can then
be represented as $y_j = C^{\rm (i)} y_j^{\rm (i)} = C^{\rm (o)} y_j^{\rm (o)}$.
The eigenvalue is 
obtained by requiring that the solutions agree at a suitable matching
point $x = \xf$, say. Thus
\begin{eqnarray}
 C^{\rm (i)} y_1^{\rm (i)} (\xf ) &=& C^{\rm (o)} y_1^{\rm (o)} (\xf ) \; , 
\nonumber \\
 C^{\rm (i)} y_2^{\rm (i)} (\xf ) &=& C^{\rm (o)} y_2^{\rm (o)} (\xf ) \; . 
\end{eqnarray}
These equations clearly have a non-trivial solution 
$(C^{\rm (i)} , C^{\rm (o)} )$
only when their determinant vanishes, 
{\ie}, when
\begin{equation}
\Delta = y_1^{\rm (i)} ( \xf ) y_2^{\rm (o)} ( \xf )
- y_2^{\rm (i)} ( \xf ) y_1^{\rm (o)} ( \xf ) = 0 \; .
\label{eq:E3.5}
\end{equation}
Equation \ref{eq:E3.5} is therefore the eigenvalue equation.

For $I$ = 4 there are two linearly independent solutions satisfying
the inner boundary conditions, and two linearly independent solutions
satisfying the outer boundary conditions. The former set $\{y_i^{\rm (i,1)} ,
y_i^{\rm (i,2)}\}$ is chosen by setting
\begin{eqnarray}
&& y_1^{\rm (i,1)} (x_1 ) = 1 \; , \qquad
  y_3^{\rm (i,1)} (x_1 ) = 0 \; , \nonumber \\
&& y_1^{\rm (i,2)} (x_1 ) = 1 \; , \qquad
  y_3^{\rm (i,2)} (x_1 ) = 1 \; , 
\label{eq:E3.6}
\end{eqnarray}
and the latter set $\{y_i^{\rm (o,1)} , y_i^{\rm (o,2)}\}$ is chosen by setting
\begin{eqnarray}
&& y_1^{\rm (o,1)} (\xs ) = 1 \; , \qquad
  y_3^{\rm (o,1)} (\xs ) = 0 \; , \nonumber \\
&& y_1^{\rm (o,2)} (\xs ) = 1 \; , \qquad
  y_3^{\rm (o,2)} (\xs ) = 1 \; . 
\label{eq:E3.7}
\end{eqnarray}
The inner and outer boundary conditions are such that, given $y_1$
and $y_3$, $y_2$ and $y_4$ may be calculated from them;
thus Eqs~\ref{eq:E3.6} and \ref{eq:E3.7}
completely specify the solutions,
which are obtained by integrating from the inner or outer boundary. 
The final solution can then be represented as 
\begin{equation}
y_j = C^{\rm (i,1)} y_j^{\rm (i,1)} + C^{\rm (i,2)} y_j^{\rm (i,2)} =
C^{\rm (o,1)} y_j^{\rm (o,1)} + C^{\rm (o,2)} y_j^{\rm (o,2)} \; .
\end{equation}
At the fitting point $\xf$ continuity of the solution
requires that
\begin{eqnarray}
&&C^{\rm (i,1)} y_j^{\rm (i,1)} (\xf ) 
+ C^{\rm (i,2)} y_j^{\rm (i,2)} (\xf ) = \\
\label{eq:E3.8}
&& C^{\rm (o,1)} y_j^{\rm (o,1)} (\xf ) + C^{\rm (o,2)} y_j^{\rm (o,2)} (\xf )
\qquad j = 1, 2, 3, 4 \; .  \nonumber
\end{eqnarray}
This set of equations only has a non-trivial solution if
\begin{equation}
\Delta = \det \left\{ \begin{array}{cccc}
y_{1,\rm f}^{\rm (i,1)}& 
y_{1,\rm f}^{\rm (i,2)}& 
y_{1,\rm f}^{\rm (o,1)}& 
y_{1,\rm f}^{\rm (o,2)} \\
\noalign{\vskip3pt}
y_{2,\rm f}^{\rm (i,1)}& 
y_{2,\rm f}^{\rm (i,2)}& 
y_{2,\rm f}^{\rm (o,1)}& 
y_{2,\rm f}^{\rm (o,2)}  \\
\noalign{\vskip3pt}
y_{3,\rm f}^{\rm (i,1)}& 
y_{3,\rm f}^{\rm (i,2)}& 
y_{3,\rm f}^{\rm (o,1)}& 
y_{3,\rm f}^{\rm (o,2)} \\
\noalign{\vskip3pt}
y_{4,\rm f}^{\rm (i,1)}& 
y_{4,\rm f}^{\rm (i,2)}& 
y_{4,\rm f}^{\rm (o,1)}& 
y_{4,\rm f}^{\rm (o,2)}
\end{array} \right\} = 0 \; ,
\label{eq:E3.9}
\end{equation}
where, {\eg},
$y_{j,\rm f}^{\rm (i,1)} \equiv y_j^{\rm (i,1)} (\xf )$.
Thus Eq.~\ref{eq:E3.9} is the eigenvalue equation in this case.

Clearly $\Delta$ as defined in either Eq.~\ref{eq:E3.5}
or Eq.~\ref{eq:E3.9} is
a smooth function of $\sigma^2$, and the eigenfrequencies are
found as the zeros of this function. This is done in the programme using
a standard secant technique.
However, the
programme also has the option for scanning through a given interval
in $\sigma^2$ to look for changes of sign of $\Delta$, possibly
iterating for the eigenfrequency at each change of sign.
Thus it is possible to search a given region 
of the spectrum completely automatically.

The programme allows the use of two different techniques for solving
the differential equations. One is the standard second-order
centred difference technique, where the differential equations are
replaced by the difference equations
\begin{equation}
{y_i^{n+1} - y_i^n \over x^{n+1} - x^n} = 
{1 \over 2 } \sum_{j=1}^I \left[ a_{ij}^n  y_j^n + a_{ij}^{n+1}  y_j^{n+1}
\right] , \quad i = 1, \ldots, I \; .
\label{eq:E3.11}
\end{equation}
Here I have introduced a mesh 
$x_1 = x^1 < x^2 < \cdots < x^N = \xs$ in $x$,
where $N$ is the total number of mesh points; 
$y_i^n \equiv y_i ( x^n )$, and 
$a_{ij}^n \equiv a_{ij} (x^n )$. These equations allow
the solution at $x = x^{n+1}$ to be determined from the
solution at $x = x^n$.

The second technique was
proposed by \citet{Gabrie1976}; 
here on each mesh interval $(x^n , x^{n+1})$ we consider the equations 
\begin{equation}
{\dd y_i^{(n)}  \over \dd x} = \sum_{j=1}^I  {\bar a}_{ij}^n  
y_j^{(n)} (x), \qquad
\hbox{\rm for } i = 1 \; , \ldots, I \; ,
\label{eq:E3.12}
\end{equation}
with constant coefficients, where 
${\bar a}_{ij}^n = 1/2 ( a_{ij}^n + a_{ij}^{n+1} )$.
These equations may
be solved analytically on the mesh intervals, and the complete
solution is obtained by continuous matching at the mesh points.
This technique clearly permits the computation of solutions varying
arbitrarily rapidly, 
{\ie}, the calculation of modes of arbitrarily high order. On the other
hand solving Eqs~\ref{eq:E3.12} involves finding the eigenvalues
and eigenvectors of the coefficient matrix, and therefore becomes
very complex and time consuming for higher-order systems. Thus in
practice it has only been implemented for systems of order 2, 
{\ie}, radial oscillations or nonradial oscillations in the Cowling
approximation.

\subsection{The relaxation technique} 

If one of the boundary conditions is dropped, the difference equations,
with the remaining boundary condition and a normalization condition,
constitute a set of linear equations for the $\{y_j^n\}$
which can be solved for any value of $\sigma$;
this set may be solved efficiently by
forward elimination and backsubstitution 
\citep[e.g.,][]{Baker1971},
with a procedure very similar to the so-called Henyey technique
\citep[e.g.,][see also Christensen-Dalsgaard 2007]{Henyey1959}
used in stellar modelling.
The eigenvalue is then found by requiring that the remaining
boundary condition, which effectively takes the role of $\Delta(\sigma)$,
be satisfied.
However, as both boundaries, at least in a complete model, are either
singular or very nearly singular, the removal of one of the
boundary conditions tends to produce solutions that are somewhat
ill-behaved, in particular for modes of high degree. This in turn
is reflected in the behaviour of $\Delta$ as a function of $\sigma$.

This problem is avoided in a variant of the relaxation
technique where the difference equations are solved separately
for $x \le \xf$ and $x \ge \xf$, by introducing a double
point $\xf^- = x^{n_{\rm f}} = x^{n_{\rm f} + 1} = \xf^+$ in the mesh.
The solution is furthermore required to
satisfy the boundary conditions \ref{eq:E3.2} and \ref{eq:E3.3},
a suitable normalization condition ({\eg}
$y_1 ( \xs ) = 1$), and continuity of all but one of the
variables at $x = \xf$, {\eg},
\begin{eqnarray}
&& y_1 ( \xf^- ) = y_1 ( \xf^+ ) \; , \nonumber \\
&& y_3 ( \xf^- ) = y_3 ( \xf^+ ) \; , \nonumber \\
&& y_4 ( \xf^- ) = y_4 ( \xf^+ ) \; , 
\end{eqnarray}
(when $I$ = 2 clearly only the first continuity condition is used)
We then set
\begin{equation}
\Delta = y_2 ( \xf^- ) - y_2 (\xf^+ ) \; ,
\label{eq:E3.15}
\end{equation}
and the eigenvalues are found as the zeros of $\Delta$, regarded as
a function of $\sigma^2$. 
With this definition, $\Delta$ may have singularities with discontinuous sign 
changes that are not associated with an eigenvalue,
and hence a little care is required in the search for eigenvalues.
However, close to an eigenvalue $\Delta$ is
generally well-behaved, and the secant iteration may be used without problems.

As implemented here the shooting technique is considerably faster than
the relaxation technique, and so should be used whenever possible
(notice that both techniques may use the difference Eqs~\ref{eq:E3.11}
and so they are numerically equivalent, in regions of the spectrum
where they both work). For {\it second-order systems\/}
the shooting technique can
probably always be used; the integrations of the inner and outer solutions
should cause no problems,
and the matching determinant in Eq.~\ref{eq:E3.5} is well-behaved.
For  {\it fourth-order systems\/}, however, this needs not be the case.
For modes where the perturbation
to the gravitational potential has little effect on the solution, the
two solutions $y_j^{\rm (i,1)}$ and $y_j^{\rm (i,2)}$, and similarly the
two solutions $y_j^{\rm (o,1)}$ and $y_j^{\rm (o,2)}$, are almost
linearly dependent, and so the matching determinant nearly vanishes for any
value of $\sigma^2$.
This is therefore the situation where the relaxation
technique may be used with advantage. 
This applies, in particular, to the calculation of modes of moderate and high
degree which are essential to helioseismology.

\subsection{Improving the frequency precision} 

To make full use of the increasingly accurate observed frequencies 
the computed frequencies should clearly at the very least match the
observational accuracy, for a given model.
Only in this way do the frequencies provide a faithful representation of
the properties of the model, in comparisons with the observations.
However, since the numerical errors in the computed frequencies are typically
highly systematic, they may affect the asteroseismic inferences even if they
are smaller than the random errors in the observations, and hence more
stringent requirements should be imposed on the computations.
Also, the fact that solar-like oscillations, and several other types of
asteroseismically interesting modes, tend to be of high radial order
complicates reaching the required precision.

The numerical techniques discussed so far are generally of second order.
This yields insufficient precision in the evaluation of the eigenfrequencies,
unless a very dense mesh is used in the computation
\citep[see also][]{Moya2007}.
The code may apply two techniques to improve the precision.

One technique \citep[cf.][]{Christ1982}
uses the fact that the frequency approximately satisfies a variational principle
\citep{Chandr1964}.%
\footnote{The variational principle is exact, formally, when the
surface Lagrangian pressure perturbation is set to zero, but not when
the match to an isothermal atmosphere is used.}
The variational expression may formally be written as
\begin{equation}
\sigma^2 = \sigma_{\rm var}^2 \equiv \Sigma(\xi)^2 
= {\CK(\xi) \over \CI(\xi)} \; ,
\label{eq:varprinc}
\end{equation}
where $\CK$ and $\CI$ are integrals over the equilibrium model depending
on the eigenfunction, here represented by $\xi$.
The variational property implies that any error $\delta \xi$ in $\xi$
induces an error in $\Sigma^2$ that is $\CO(|\delta\xi|^2)$.
Thus by substituting the computed eigenfunction into the variational expression
a more precise determination of $\sigma^2$ should result.
This has indeed been confirmed \citep{Christ1982, Christ1991, Christ1994}.

The second technique uses explicitly that 
the difference scheme \ref{eq:E3.11}, which is used by one
version of the shooting technique, and the relaxation technique,
is of second order.
Consequently the truncation errors in the eigenfrequency and eigenfunction
scale as $N^{-2}$.
If $\sigma ( N/2 )$ and $\sigma (N)$ are the eigenfrequencies obtained
from solutions with $N/2 $ and $N$ meshpoints, the leading-order
error term therefore cancels in 
\begin{equation}
\sigma_{\rm Ri} = {1 \over 3} [ 4 \sigma (N) - \sigma ( {1\over 2} N ) ] \; .
\label{eq:E3.18}
\end{equation}
This procedure, known as {\it Richardson extrapolation},
was used by 
\citet{Shibah1981}.
It provides an estimate of the eigenfrequency that is substantially
more accurate than $\sigma ( N )$, although of course at some
added computational expense.
Indeed, since the error in the representation \ref{eq:E3.11} depends only
on even powers of $N^{-1}$, the leading term of the error in 
$\sigma_{\rm Ri}$ is $\CO(N^{-4})$.

Even with these techniques the precision of the computed frequencies 
may be inadequate if the mesh used in stellar-evolution calculations is
used also for the computation of the oscillations.
The number of meshpoints is typically relatively modest and the distribution
may not reflect the requirement to resolve properly the eigenfunctions of
the modes.
\citet{Christ1991} discussed techniques to redistribute the mesh in a way
that takes into account the asymptotic behaviour of the eigenfunctions;
a code to do so, based on four-point Lagrangian interpolation,
is included in the ADIPLS distribution package.
On the other hand, for computing low-order modes (as are typically relevant
for, say, $\delta$ Scuti or $\beta$ Cephei stars), the original mesh of
the evolution calculation may be adequate.

It is difficult to provide general recommendations concerning the required
number of points or the need for redistribution,
since this depends strongly on the types of modes and the properties of
the stellar model.
It is recommended to carry out experiments varying the number and
distribution of points to obtain estimates of the intrinsic precision
of the computation \citep[e.g.,][]{Christ1991, Christ1994}.
In the latter case, considering simple polytropic models,
it was found that 4801 points yielded a relative precision substantially
better than $10^{-6}$ for high-order p modes, when Richardson extrapolation 
was used.

In the discussion of the frequency calculation
it is important to distinguish between {\it precision\/} 
and {\it accuracy\/}, the latter obviously referring to the extent to
which the computed frequencies represent what might be considered the `true'
frequencies of the model.
In particular, the manipulations required to derive Eq.~\ref{eq:varprinc}
and to demonstrate its variational property depend on the equation of
hydrostatic support being satisfied.
If this is not the case, as might well happen in an insufficiently careful
stellar model calculation, the value determined from 
the variational principle may be quite precise, in the sense of numerically
stable, but still unacceptably far from the correct value.
Indeed, a comparison between $\sigma_{\rm var}$ and $\sigma_{\rm Ri}$
provides some measure of the reliability of the computed frequencies
\citep[e.g.][]{Christ1991}.

\section{Computed quantities}
\label{sec:results}

The programme finds the order of the mode according to the 
definition developed by \citet{Scufla1974} and \citet{Osaki1975},
based on earlier work by \citet{Eckart1960}.
Specifically, the order is defined by
\begin{equation}
n =
- \sum_{x_{z1}> 0} \sign \left( y_2  {\dd y_1 \over \dd x }\right) + n_0 \; .
\label{eq:E4.1}
\end{equation}
Here the sum is over the zeros $\{x_{z1}\}$ in $y_1$ (excluding the
centre), and $\sign$ is the sign function, $\sign(z) = 1$ if $z > 0$
and $\sign(z) = -1$ if $z < 0$.
For a complete model that includes the centre
$n_0 = 1$ for radial oscillations
and $n_0 = 0$ for nonradial oscillations. Thus the lowest-order
radial oscillation has order $n = 1$. Although this is contrary to the
commonly used convention of assigning order 0 to the fundamental radial
oscillation, the convention used here is in fact the more reasonable,
in the sense that it ensures that $n$ is invariant under a continuous
variation of $l$ from 0 to 1. With this definition $n > 0$ for p modes, 
$n = 0$ for f modes, and $n < 0$ for g modes, at least in simple models.

It has been found that this procedure
has serious problems for dipolar modes in centrally condensed models
\citep[e.g.,][]{Lee1985, Guenth1991, Christ1994}.
The eigenfunctions $y_1$ are shifted such that nodes disappear or otherwise
provide spurious results when Eq.~\ref{eq:E4.1} is used to determine the
mode order.
A procedure that does not suffer from this difficulty has recently been
developed by \citet{Takata2006b};
I discuss it further in Section~\ref{sec:develop}.

A powerful measure of the characteristics of a mode is provided by the
{\it normalized inertia\/}.
The code calculates this as 
\begin{eqnarray}
\hat E &=& {\int_{r_1}^{R_{\rm s}} [ \xi_r^2 + l ( l + 1) \xih^2 ]
 \rho  r^2  \dd r \over
  M [ \xi_r (R_{\rm phot} )^2 + l(l+1) \xih (R_{\rm phot} )^2]} \nonumber \\
  &=& { \int_{x_1}^{\xs} \left[ y_1^2
  + y_2^2  / l ( l + 1) \right]  q U \dd x / x
  \over 4 \pi [ y_1 (x_{\rm phot} )^2 + y_2 (x_{\rm phot} )^2/l(l+1)] } \; .
\end{eqnarray}
(For radial modes the terms in $y_2$ are not included.)
Here $r_1 =  R  x_1$ and $R_{\rm s} = R  \xs$ are the
distance of the innermost mesh point from the centre and the surface radius,
respectively, 
and $x_{\rm phot} = R_{\rm phot}/R =1$ is the fractional photospheric radius.
The normalization at the photosphere is to some extent arbitrary, of course,
but reflects the fact that many radial-velocity observations use lines 
formed relatively deep in the atmosphere.
A more common definition of the inertia is
\begin{equation}
E = 4 \pi \hat E = {M_{\rm mode} \over M} \; ,
\end{equation}
where $M_{\rm mode}$ is the so-called {\it mode mass}.

The code has the option to output the eigenfunctions, in the form of
$\{y_j(x^n)\}$.
In addition (or instead) the displacement eigenfunctions can be output
in a form indicating the region where the mode predominantly resides,
in an energetical sense, as
\begin{eqnarray}
z_1 (x) &=& \left( {4 \pi r^3 \rho  \over M }\right)^{1/2}  y_1 (x)
= \left( {4 \pi r^3 \rho  \over M }\right)^{1/2}  {\xi_r(r) \over R} \; ,
\nonumber \\
z_2 (x) &=& {1 \over \sqrt {l ( l + 1) } }
\left( {4 \pi r^3 \rho  \over M }\right)^{1/2} y_2 (x) \nonumber \\
&=& \sqrt {l ( l + 1) }
  \left( {4 \pi r^3 \rho  \over M }\right)^{1/2} { \xih(r) \over R} \; 
\end{eqnarray}
(for radial modes only $z_1$ is found).
These are defined in such a way that
\begin{equation}
\hat E = {\int_{x_1}^{\xs} [ z_1^2 + z_2^2 ]  \dd x / x  
  \over 4 \pi [ y_1 (x_{\rm phot} )^2 + y_2 (x_{\rm phot} )^2/l(l+1)] } \; .
\label{eq:E4.4}
\end{equation}

The form provided by the $z_i$ is also convenient, e.g., for computing
rotational splittings $\delta \omega_{n l m} = \omega_{n l m} - \omega_{n l 0}$
\citep[e.g.,][]{Gough1981}, where $\omega_{n l m}$ is the frequency of
a mode of radial order $n$, degree $l$ and azimuthal order $m$.
For slow rotation the splittings are obtained from first-order perturbation
analysis as
\begin{equation}
\delta \omega_{n l m} = m \int_0^{R_{\rm s}} \int_0^\pi
K_{n l m} (r, \theta) \Omega(r , \theta) r \dd r \dd \theta \; ,
\label{eq:rotsplit}
\end{equation}
characterized by {\it kernels\/} $K_{n l m}$,
where in general the angular velocity $\Omega$ depends on both $r$ and
$\theta$.
The code has built in the option to compute kernels 
for first-order rotational splitting in the special case where
$\Omega$ depends only on $r$.

\section{Further developments}
\label{sec:develop}

Several revisions of the code have been implemented in preliminary form or
are under development.
A substantial improvement in the numerical solution of the oscillation
equations, particularly for high-order modes, is the installation of
a fourth-order integration scheme, based on the algorithm of \citet{Cash1980}.
This is essentially operational but has so far not been carefully tested.
Comparisons with the results of the variational expression and
the use of Richardson extrapolation, of the same formal order, will
be particularly interesting.

As discussed by \citet{Moya2007} the use of $p'$ 
(or, as here, $\xih$) as one of the integration variables has the
disadvantage that the quantity $A$ enters into the oscillation equations.
In models with a density discontinuity, such as results if the model
has a growing convective core and diffusion is neglected,
$A$ has a delta-function singularity at the point of the discontinuity.
In the ADIPLS calculations this is dealt with by replacing the discontinuity
by a very steep and well-resolved slope.
However, it would obviously be an advantage to avoid this problem altogether.
This can be achieved by using instead the Lagrangian pressure perturbation
$\delta p$ as one of the variables.
Implementing this option would be a relatively straightforward modification
to the code and is under consideration.

The proper classification of dipolar modes of low order in centrally
condensed models has been a long-standing problem in the theory of
stellar pulsations, as discussed in Section~\ref{sec:results}.
Such a scheme must provide a unique order for each mode, such that the
order is invariant under continuous changes of the equilibrium model,
e.g., as a result of stellar evolution.
As a major breakthrough, Takata in a series of papers has elucidated
important properties of these modes and defined a new classification scheme
satisfying this requirement \citep{Takata2005, Takata2006a, Takata2006b}.
A preliminary version of this scheme has been implemented and tested;
however, the latest and most convenient form of the Takata classification
still needs to be installed.

A version of the code has been established which computes the first-order
rotational splitting for a given rotation profile $\Omega(r)$, 
in addition to setting up the corresponding kernels.
This is being extended by K. Burke, Sheffield, to cover also
second-order effects of rotation, based on the formalism of \citet{Gough1990}.
An important motivation for this is the integration, discussed by
\citet{Christ2007},
of the pulsation calculation with the
ASTEC evolution code to allow full calculation of oscillation frequencies
for a model of specified parameters (mass, age, initial rotation rate, etc.)
as the result of a single subroutine call.

\begin{acknowledgements}
I am very grateful to W. Dziembowski and D.~O.~Gough for illuminating
discussions of the properties of stellar oscillations,
and to A. Moya and M. J. P. F. G. Monteiro for organizing the comparisons
of stellar oscillation and model calculations within the ESTA collaboration.
I thank the referee for useful comments which, I hope, have helped 
improving the presentation.
This project is being supported by
the Danish Natural Science Research Council and by
the European Helio- and Asteroseismology Network (HELAS),
a major international collaboration funded by the European Commission's
Sixth Framework Programme.
\end{acknowledgements}

%

\end{document}